\documentclass[a4paper,11pt]{article}

\usepackage{jheppub} 

\usepackage[T1]{fontenc} 

\usepackage{color}

\def\Eq#1{\begin{equation} #1 \end{equation}}
\def\Eqr#1{\begin{eqnarray} #1 \end{eqnarray}}
\def\Eqrsubl#1#2{\begin{subequations}\label{#1}\Eqr{#2}\end{subequations}}

\newcommand{\nn}{\nonumber}

\newcommand{\bea}{\begin{eqnarray}}
\newcommand{\eea}{\end{eqnarray}}

\def\Xsp{{\rm X}}

\def\Ysp{{\rm Y}}

\def\X5sp{{\rm X}_5}
\def\Y3sp{{\rm Y}_3}
\def\Z3sp{{\rm Z}_3}

\def\lap{{\triangle}}

\title{\boldmath Fractional black $p$-branes on orbifold 
${\mathbb C}^n/{\mathbb Z}_n$}

\author[a,b]{Muneto Nitta}
\author[b,c]{Kunihito Uzawa}

\affiliation[a]{Department of Physics, Keio University,\\ 
Hiyoshi 4-1-1, Yokohama, Kanagawa 223-8521, Japan}
\affiliation[b]{Research and Education 
Center for Natural Sciences, Keio University,\\ 
Hiyoshi 4-1-1, Yokohama, Kanagawa 223-8521, Japan}
\affiliation[c]{Department of Physics, School of Science 
and Technology, Kwansei Gakuin University,\\
 Sanda, Hyogo 669-1337, Japan}

\abstract{The recent discovery of an explicit solution of 
a black hole on the resolved orbifold 
${\mathbb C}^n/{\mathbb Z}_n$
makes it possible to investigate the existence  
of $p$-branes on the orbifold. In particular, 
it is possible with 
reasonable precision to verify the prediction that an M2-brane 
on ${\mathbb C}^4/{\mathbb Z}_4$ in eleven dimensions 
and a D3-brane on ${\mathbb C}^3/{\mathbb Z}_3$ in ten dimensions 
have a 
family of black $p$-branes on the orbifold 
${\mathbb C}^n/{\mathbb Z}_n$. 
These solutions are extremal 
and 
have regular horizons $S^{2n-1}/{\mathbb Z}_n$ 
without any naked singularity, 
with near horizon geometries 
 AdS${}_{p+2}\times S^{2n-1}/{\mathbb Z}_n$.}

\begin{document} 
\maketitle
\flushbottom


\section{Introduction}
\label{sec:introduction}

In recent developments involving string theory and 
supergravities, 
$p$-branes carrying higher-dimensional charges 
have played crucial roles to understand their 
non-perturbative dynamics 
\cite{Gibbons:1987ps, Dabholkar:1989jt, Dabholkar:1990yf, 
Callan:1991ky, Horowitz:1991cd, Stelle:1996tz, Duff:1996hp, 
Argurio:1998cp}. 
The terminology ``$p$-brane'' generally should be used to indicate 
a classical solution which is extended in $p$ directions 
in the context of a gravity theory. $p$-brane solutions 
have $p$ spacelike translational Killing vectors 
and carry the charge of an antisymmetric tensor field. 
Since they have the mass saturating a lower bound 
given by the charge, they are called extremal, 
which are also called as 
Bogomol'nyi-Prasad-Sommerfield (BPS) states 
\cite{Bogomolny:1975de, Prasad:1975kr, 
Coleman:1976uk}.  
Four-dimensional charged black holes are 
prototypes of $p$-branes for the case 
of $p=0$. The $p$-branes are also called solitons 
in the sense that they are classical solutions. 
In many instances it
is important to know about black hole solutions 
of such objects. 
For example, a class of extended black hole solutions 
can be found simply by taking 
the product of $(D-d)$-dimensional flat space and 
the $d$-dimensional Schwarzschild solution. 
Since these spacetimes are Ricci flat, we will 
solve the field equations of low-energy effective 
theory  of string theory.  
They give a $p\,(=D-d)$ brane 
surrounded by an event horizon, which is a black 
$p$-brane. These solutions are described by 
a physical parameter which can be interpreted as 
the tension, {\it i.e.}, the mass per unit 
$p$-volumes of the $p$-branes. 

The $p$-branes can be put in more general manifolds. 
In particular, they gain a lot of interesting properties 
when these objects are located at an orbifold singularity 
\cite{Douglas:1996sw, Douglas:1997de}, 
such as a D3-brane on ${\mathbb C}^3/{\mathbb Z}_3$ 
in type-II string theory 
\cite{Gukov:1998kn,Oh:2002sv,Krishnan:2008kv}
and an M2-brane on ${\mathbb C}^4/{\mathbb Z}_4$ 
in M-theory 
explicitly constructed in 
Refs.~\cite{Singh:2008ix,Krishnan:2009nw}. 
One of characteristic features is that
at the orbifold singularities, the mass and charge of the $p$-branes
become fractional, and thus they 
cannot move from the orbifold singularity 
freely due to the Dirac's quantization condition.  
Such fractional charges of solitons stucked at 
orbifold singularities are a common feature among 
Yang-Mills instantons \cite{Nakajima,Nakajima2}, 
vortices \cite{Kimura:2011wh}, 
and D-branes 
\cite{Douglas:1996sw,Douglas:1997de,Eto:2004vy} 
on orbifold singularities.
Another interesting property is that 
orbifold singularities are resolved 
in the D-brane world volume theory \cite{Douglas:1996sw,Douglas:1997de}. 
Thus, one can expect that 
the spacetime itself 
becomes regular without 
any naked singularity once $p$-branes are placed 
at the orbifold singularities. 
This is important when one finds 
hints that the gravity theory giving 
a $p$-brane on the orbifold 
is actually an already known and well-defined supergravity, 
thus giving a handle on the strong coupling dynamics of 
string theory.

An example of a theory for which a cosmological model  
such as a brane world scenario 
\cite{Randall:1999ee, Randall:1999vf} 
was found owing to the properties of the orbifold geometry 
is the strongly coupled ${\rm E}_8\times {\rm E}_8$ 
heterotic string theory in ten dimensions. This is equivalent 
to the eleven-dimensional limit of M-theory compactified on 
an $S^1/{\mathbb Z}_2$ orbifold. A set of ${\rm E}_8$ gauge 
fields is located at each ten-dimensional orbifold fixed plane. 
The concrete solution in five-dimensional theory was later 
constructed by an orbifold compactification 
\cite{Lukas:1998yy, Lukas:1998tt}. 
The idea that our universe may be a 3-brane in a 
higher dimensional spacetime gives also a supersymmetric 
realization of the brane-world in type IIB supergravity 
\cite{Duff:2000az, Cvetic:2000gj}. 

One of the most interesting issues of this kind, and the main 
focus in this paper 
concern $p$-branes with vanishing scalar fields 
in $D$ dimensions.
The eleven-dimensional supergravity theory is 
for instance believed to have a gravity and four-form 
field strength for the bosonic sector 
if the higher-order terms are negligible. 
One can characterize this by saying that it has a 
charged black hole (M-brane) 
without any naked singularity in a neighborhood and 
everywhere outside of the event horizon 
\cite{Stelle:1998xg}. The physical mass 
and charge for the black hole of these M-branes has been 
worked out in detail \cite{Lu:1993vt, Argurio:1998cp}. 
The mass formula is determined by the 
Arnowitt-Deser-Misner (ADM) formalism 
\cite{Arnowitt:1960es,Arnowitt:1962hi,Lu:1993vt}  
and so it will be enough in this paper to 
find a $p$-brane solution.

What makes these questions accessible for study is that an 
explicit and amazingly simple description of a black hole 
on the resolved orbifold 
 ${\mathbb C}^n/{\mathbb Z}_n$ 
 has recently been given in the Einstein-Maxwell 
system \cite{Nitta:2020pzo} 
as a generalization of $D=5$ black hole on the Eguchi-Hanson space 
\cite{Ishihara:2006pb,Ishihara:2006iv}
(see Refs.~\cite{Tatsuoka:2011tx,Dehghani:2005zm,
Dehghani:2006aa} for another higher dimensional generalization). 
 The basic two-form field strength 
has a 1-form gauge potential since it was seen 
in Ref.~\cite{Nitta:2020pzo} 
to carry a point charge, but on backgrounds does not give 
any naked singularity. 
Underlying geometry of the resolved orbifold 
 ${\mathbb C}^n/{\mathbb Z}_n$ is to formulate this manifold 
 as a complex line bundle over 
${\mathbb C}{\rm P}^{n-1}$ admiting a Ricci-flat K\"{a}hler metric 
\cite{Higashijima:2001vk,Higashijima:2001fp,Higashijima:2002px}.

In this paper, we construct black $p$-brane solutions on 
the resolved orbifold 
 ${\mathbb C}^n/{\mathbb Z}_n$ in 
 a $D=2n+p+1$ dimensional ($p+1$)-form Einstein system. 
 The mass and charge are proportional to each other 
 and the solution is extremal 
 or BPS. 
 The near horizon geometry is 
 AdS${}_{p+2}\times S^{2n-1}/{\mathbb Z}_n$ space.  
The case of $p=0$, $D=5$ ($n=2$) reduces to 
a black hole on the Eguchi-Hanson space 
\cite{Ishihara:2006pb}, 
while the one of $p=0$ and $D=2n+1$ 
is for a black hole on the orbifold 
${\mathbb C}^n/{\mathbb Z}_n$ \cite{Nitta:2020pzo}.
As previously known $p$-brane solutions, 
our solutions reduce in
the case of $p=3$ and $D=10$ 
to a D3-brane on ${\mathbb C}^3/{\mathbb Z}_3$ 
in type-II string theory 
with a near horizon geometry 
AdS${}_{5}\times S^{5}/{\mathbb Z}_3$ 
\cite{Gukov:1998kn,Oh:2002sv,Krishnan:2008kv},
and 
in the case of $p=2$ and $D=11$ 
to an M2-brane on ${\mathbb C}^4/{\mathbb Z}_4$ 
in M-theory  
with a near horizon geometry 
AdS${}_{4}\times S^{7}/{\mathbb Z}_4$ 
  \cite{Singh:2008ix,Krishnan:2009nw}. 
However, our solutions with general $p$ and $D$ 
do not rely on any string theory origin, 
and thus have a potential application to 
more general cases such as brane-world scenarios.

This paper is organized as follows.
In section \ref{sec:p}, we show that simple field equations 
exist as an almost immediate consequence of the ansatz for fields 
we will impose. 
To see the $p$-brane on the orbifold ${\mathbb C}^n/{\mathbb Z}_n$ 
with $n\ge 1$ requires a more elaborate construction to which we 
then turn in section \ref{cpn}. In the process, 
$(p+2)$-dimensional AdS spacetime AdS${}_{p+2}$
makes an expected appearance near horizon limit of the 
$D$-dimensional $p$-brane background. 
A reasonably compelling argument for the existence 
of these vacua will be presented. 
We then go on in section \ref{sec:discussions} to summarize 
our results and comments about some of the other $p$-branes.

\section{Classical $p$-brane solutions}
\label{sec:p}
We now go on to the formulation of finding 
$p$-brane solutions. In this section, we will present a 
general action in $D$ dimensions. 
Here, we consider a theory including gravity, 
scalar field, and antisymmetric tensor
fields of arbitrary rank 
$(p+1)$  with its field strength of rank $(p+2)$\,. 
Then, the most general action is the following 
\cite{Duff:1994an, Argurio:1998cp}
\Eq{
S=\frac{1}{2\kappa^2}\int d^Dx\sqrt{-g}\left[R
 -\frac{1}{2\cdot \left(p+2\right)!}
 \,F_{(p+2)}^2\right],
\label{p:action:Eq}
}
where $\kappa^2$ states again the $D$-dimensional 
gravitational constant. 
The reduction to the cases expressed in Eq.~(\ref{p:action:Eq}) 
is straightforward. If we consider $D=11$ supergravity, we 
only have a 4-form field strength\,. 

In this paper, we focus on a particular case of the action 
(\ref{p:action:Eq}), where there is a single $(p+2)$-form 
field strength. In this setup, we can impose several 
symmetries in the background, which will allow us to constrain 
the fields in such a way that it will be possible to obtain a 
solution to the field equations derived from the action. 
The solutions found in this procedure will then give 
useful methods when we construct more general solutions 
in issues with less symmetries. It is also useful to consider 
the configurations with several branes like the multiply 
charged solutions.  
The matter fields presented in this paper is 
well-known as giving the single brane solutions 
\cite{Gibbons:1987ps, Horowitz:1991cd, Duff:1996hp, 
Argurio:1998cp}. 

Now we can write the equations of motion and Bianchi identities 
for this simpler case, containing only gravity, 
a $(p+2)$-form field strength and the scalar field. 
The equations of motion from Eq.~(\ref{p:action:Eq}) can be  
expressed as  
\Eqrsubl{p:fe:Eq}{
&&\hspace{-1cm}R_{MN}=\frac{1}{2\cdot \left(p+2\right)!}\,
\left[(p+2)F_{MA\cdots B} {F_N}^{A\cdots B}
-\frac{p+1}{D-2}g_{MN} F^2_{(p+2)}\right],
   \label{p:Einstein:Eq}\\
&&\hspace{-1cm}d\left[\ast F_{(p+2)}\right]=0\,,
   \label{p:gauge:Eq}
}
where $\ast$ denotes the Hodge dual operator in the
$D$-dimensional spacetime\,. 
The $(p+2)$-form field strength $F_{(p+2)}$ 
coming from $(p+1)$-form gauge potential $A_{(p+1)}$ has to 
obey the Bianchi identities 
\Eq{
dF_{(p+2)}=0\,.
 \label{p:bi:Eq}
} 
The three equations (\ref{p:fe:Eq}), (\ref{p:bi:Eq}), 
are the ones which we will use in this section to find 
the $p$-brane solutions. 

We now impose the symmetries of the background 
in order to simplify the field equations 
by restricting us to particular field configurations. 
A $p$-brane solution living in a $D$-dimensional spacetime 
is in general specified by the fact that $p$ spacelike directions 
can be described longitudinal to the $p$-brane while
the remaining $D-p-1$ spacelike direction are considered 
transverse to the brane. 
When we consider single $p$-brane solutions 
carrying a single charge, it is natural to assume that 
$p$ longitudinal directions are all equivalent. 
Although the timelike direction can be also longitudinal 
to the world-volume of the $p$-brane, it will not be 
considered as equivalent to the other longitudinal
directions in the general case.

Since the $p$-brane is a uniform object,  
it should not single out particular points or 
regions of its volume. Then, it has the $p$ spacelike 
longitudinal directions which is defined as many 
translational invariant directions of the solution. 
For a static $p$-brane, as we will discuss 
in this paper, there is another translationally  
invariant direction which is the timelike one. 
Since the $p$-brane is localized at a point in the 
transverse space, invariance under translations is 
thus broken \cite{Argurio:1998cp}. 
If we assume that the $p$-brane is static with 
vanishing angular momentum, we can postulate 
spherical symmetry in the transverse space.  

In this paper, the further restrictions that we will 
implement on the fields will specify the solution 
to be in a class, which is extremal $p$-brane 
because field equations are easily solved, and also these 
extremal $p$-branes turn out to be physically interesting, 
being identified in some cases to fundamental objects in 
the string theory as well as the general relativity. 

Extremal $p$-brane roughly corresponds to the 
description that the mass of the $p$-brane is equal 
or proportional to its charge. 
This states that the $p$-brane is fully characterized 
 by only the $p$-brane charge in the 
absence of angular momenta. 
From the viewpoint of the worldvolume of the $p$-brane, 
it is seen as a configuration carrying no energy at all 
\cite{Tseytlin:1996hi}. 
It seems thus to describe this configuration as 
the flat space. 
If there is any mass excitation, or any 
departure from extremality in $D$-dimensional theory, 
the Lorentz invariance is broken  
in $\left(p+1\right)$-dimensional worldvolume spacetime 
\cite{Argurio:1998cp}. 

Now we set a metric of extremal $p$-brane in $D$ dimensions 
\Eq{
ds^2=h^a(y)q_{\mu\nu}(x)dx^\mu dx^\nu+h^b(y)
u_{ij}(y)dy^idy^j\,,
   \label{p:metric:Eq}
}
where we split the coordinates accordingly 
in two sets, ${x^\mu}$ and ${y^i}$ with $\mu=0\,,~
\cdots\,,~p$ and $i=1\,,~\cdots\,,~
D-p-1$\,, and $q_{\mu\nu}(x)$ and $u_{ij}(y)$ 
are metrics of $(p+1)$-dimensional specetime, 
$(D-p-1)$-dimensional space depending only on 
$x^\mu$\, and $y^i$\,, respectively. 
We assume that the constants $a$ and $b$ are 
given by 
\Eq{
a=-\frac{2}{p+1}\,,~~~~b=\frac{2}{D-p-3}\,.
} 
The coordinates $x^\mu$ span the directions 
longitudinal to the brane, and the $y^i$ states 
the coordinates of the transverse space. We have 
chosen the timelike direction $x^0=t$\,. 
The function $h$ depends only on the coordinates 
of the transverse space to the $p$-brane. 

We go on characterizing which components of the 
$(p+2)$-form field strength are relevant to the 
$p$-brane solutions. 
In this paper, we try to formulate the following 
ansatz for a $(p+2)$-form field strength. 
A $(p+2)$-form satisfies 
an electric ansatz if it is of the form
\Eq{
F_{(p+2)}=\sqrt{\frac{2\left(D-2\right)}
{\left(p+1\right)\left(D-p-3\right)}}\,d\left(h^{-1}\right)\wedge 
\Omega\left(\Xsp\right)\,,
   \label{p:F:Eq}
}
where ``electric'' means the ansatz (\ref{p:F:Eq}) 
solves trivially its Bianchi identities 
(\ref{p:bi:Eq}), and we have defined
\Eq{
\Omega\left(\Xsp\right)=\sqrt{-q}\,dx^0\wedge 
dx^1\wedge\cdots\wedge dx^p\,.
}
Here, $q$ states the determinant of the 
$(p+1)$-dimensional metric and X denotes the 
$(p+1)$-dimensional spacetime which is longitudinal to 
the world-volume of the brane. 
It is straightforward to find that with 
such an ansatz the Bianchi identities 
(\ref{p:bi:Eq}) are trivially satisfied. 
Taking into account the metric (\ref{p:metric:Eq}), 
the field equation (\ref{p:gauge:Eq}) 
for the antisymmetric tensor becomes 
\Eq{
\lap_\Ysp h=0\,,
   \label{p:f2:Eq}
}
where $\lap_\Ysp$ states the Laplace operator constructed 
from the metric $u_{ij}(y)$\,, and Y is 
the $(D-p-1)$-dimensional space transverse to the $p$-brane. 

We are now ready to collect all the above results 
and to express the Einstein equations for the metric 
(\ref{p:metric:Eq}) and the $(p+2)$-form field 
strength (\ref{p:F:Eq}). The Einstein equations are 
totally diagonal form in these coordinates, 
and we write here the equations in the following:
\Eqrsubl{p:Ein:Eq}{
&&R_{\mu\nu}(\Xsp)+\frac{1}{p+1}q_{\mu\nu}(x)
h^{-\frac{2(D-2)}{(p+1)(D-p-3)}-1}
\lap_\Ysp h=0\,,\\
&&R_{ij}(\Ysp)-\frac{1}{D-p-3}u_{ij}(y)
h^{-1}\lap_\Ysp h=0\,,
}
where $R_{\mu\nu}(\Xsp)$ and $R_{ij}(\Ysp)$ 
denote the Ricci tensor on X spacetime and Y space 
with respect to 
the metrics $q_{\mu\nu}(x)$ and 
$u_{ij}(y)$\,, 
respectively\,.
If we require that the Ricci tensor $R_{\mu\nu}(\Xsp)$ 
depends only on 
the coordinates $x^\mu$\,, these equations can be 
reduced to 
\Eq{
R_{\mu\nu}(\Xsp)=0\,,~~~~~R_{ij}(\Ysp)=0\,,~~~~~
\lap_\Ysp h=0\,.
   \label{p:fields:Eq}
}

To summarize, for any $h$ of the form $h=h(y)$\,, 
and $(p+2)$-form $F_{(p+2)}$ on Y space satisfying
$\lap_\Ysp h=0$\,, the metric (\ref{p:metric:Eq}) with 
$R_{\mu\nu}(\Xsp)=0$\,, $R_{ij}(\Ysp)=0$\,, and the 
field strength given by (\ref{p:F:Eq}) 
yield a solution to the $D$-dimensional theory 
\cite{Duff:1994an, Duff:1996hp, 
Stelle:1996tz, Stelle:1998xg, Argurio:1998cp}. 

\section{Black $p$-branes 
on the orbifold ${\mathbb C}^{n}/{\mathbb Z}_{n}$ }
\label{cpn}

The solution of field equation is thus fully characterized by 
the equation $\lap_\Ysp h=0$\,.
Now we show that the a solution to 
Eq.~(\ref{p:fields:Eq}) has a complex line bundle 
over the complex projective space $\mathbb{C}{\rm P}^{n-1}$\,. 
This is a generalization of the solution discussed in 
Refs.~\cite{Ishihara:2006pb, Nitta:2020pzo}. 

\subsection{Black $p$-brane solution}
Let us consider the metric $u_{ij}(y)$ 
on the $(2n)$-dimensional space Y in 
Eq.~(\ref{p:metric:Eq})
\Eq{
u_{ij}(y)dy^idy^j=
dr^2+r^2\left[\left\{d\rho+\sin^2\xi_{n-1}
\left(d\psi_{n-1}+\frac{1}{2(n-1)}\omega_n\right)
\right\}^2+ds^2_{\mathbb{C}{\rm P}^{n-1}}\right].
}
Here $r$ is a radial coordinate, 
$\rho$ is a coordinate of $S^1$, 
$\xi_{n-1}$ and $\psi_{n-1}$ are coordinates of 
the $\mathbb{C}{\rm P}^{n-1}$ space 
with the ranges 
$0\le\xi_{n-1}\le \pi/2$\,,~
$0\le \psi_{n-1}\le 2\pi$\,. 
$\omega_{n-1}$ and 
$ds^2_{\mathbb{C}{\rm P}^{n-1}}$ state 
a one-form and a metric on the  
${\mathbb{C}}{\rm P}^{n-1}$ space, respectively,  
recursively defined as
\cite{Dehghani:2005zm, Dehghani:2006aa, Tatsuoka:2011tx}
\Eqr{
ds^2_{\mathbb{C}{\rm P}^{n-1}}&=&
2n\left[d\xi_{n-1}^2+
\sin^2\xi_{n-1}\,\cos^2\xi_{n-1}
\left\{d\psi_{n-1}+\frac{1}{2(n-1)}
\omega_{n-2}\right\}^2\right.\nn\\
&&\left.+\frac{1}{2(n-1)}\sin^2\xi_n\,
ds^2_{\mathbb{C}{\rm P}^{n-2}}
\right],
} 
and 
\Eqrsubl{cpn:cp1:Eq}{
\omega_{n-2}&=&2(n-1)\sin^2\xi_{n-2}
\left[d\psi_{n-2}+\frac{1}{2(n-2)}\omega_{n-3}
\right],\\
ds^2_{\mathbb{C}{\rm P}^1}&=&4\left(
d\xi_1^2+\sin^2\xi_1\,\cos^2\xi_1\,d\psi_1^2\right)\,,\\
\omega_1&=&4\sin^2\xi_1\,d\psi_1\,,
}
where $(r, \rho)$ describes a complex line,
and $\rho$ together with ${\mathbb C}{\rm P}^n$ 
denote a $(2n-1)$-dimensional sphere 
$S^{2n-1}/{\mathbb Z}_n = S^{D-p-2}/{\mathbb Z}_n $\,, 
which is actually an event horizon.

If we set $h=h(r)$\,, the equation $\lap_\Ysp h=0$ gives 
the solution 
\Eq{
h(r)=c_0+\frac{c_1}{r^{D-p-3}}\,,
}
where $c_0$ and $c_1$ denote constants. 
The $D$-dimensional metric then writes 
\Eq{
ds^2=\left(c_0+\frac{c_1}{r^{D-p-3}}\right)^{-\frac{2}{p+1}}
q_{\mu\nu}(x)dx^\mu dx^\nu
+\left(c_0+\frac{c_1}{r^{D-p-3}}
\right)^{\frac{2}{D-p-3}}u_{ij}(y)dy^i\,dy^j\,.
\label{cpn:metric:Eq}
}
The solution (\ref{cpn:metric:Eq}) has the same form as 
a standard non-dilatonic $p$-brane
solution with Ricci flat transverse space 
\cite{Stelle:1996tz, Argurio:1998cp, Gibbons:1994vm} due to 
the structure of the field equations (\ref{p:fields:Eq}). 

As a particular example, we consider the case 
$q_{\mu\nu}=\eta_{\mu\nu}$
where $\eta_{\mu\nu}$ is the 
$(p+1)$-dimensional Minkowski metric. 
Now we further define a new coordinate $\bar{r}$ by 
$\bar{r}=r^\alpha$\,, where $\alpha$ is expressed as 
\Eq{
\alpha=\frac{D-p-3}{p+1}\,.
}
Keeping the values of these coordinates finite, 
the metric in the limit $\bar{r}\rightarrow 0$ then becomes
\Eqr{
ds^2&=&c_1^{-\frac{2}{p+1}}\,\bar{r}^2\,
\eta_{\mu\nu}dx^\mu dx^\nu
+\left(\frac{c_1^{\frac{1}{D-p-3}}}{\alpha}\right)^2
\,\frac{d\bar{r}^2}{\bar{r}^2}\nn\\
&&+c_1^{\frac{2}{D-p-3}}
\left[\left\{d\rho+\sin^2\xi_{n-1}
\left(d\psi_{n-1}+\frac{1}{2(n-1)}\omega_n\right)
\right\}^2+ds^2_{\mathbb{C}{\rm P}^{n-1}}\right].
}
Hence, $D$-dimensional metric becomes an 
AdS${}_{p+2}\times S^{D-p-2}/{\mathbb Z}_n =$  
AdS${}_{p+2}\times S^{2n-1}/{\mathbb Z}_n$ space. 

A $p$-brane solution without any scalar field near the horizon 
describes AdS space is a consequence of the fact that 
if the $r\rightarrow 0$ limit is taken (also called the 
near horizon limit), the contribution of field strength becomes 
strong and the spacetime is close to the Freund-Rubin type 
of compactification \cite{Freund:1980xh} 
with a $(p+2)$-form gauge field strength 
while it will vanish in the limit $r\rightarrow\infty$\,. 
Then, it is possible then to write a theory of these ``low-energy'' 
modes, and the $(p+2)$-dimensional Minkowski spacetime is 
recovered in the asymptotic limit.

The case of $p=0$, $D=5$ reduces to 
a black hole on the Eguchi-Hanson space 
\cite{Ishihara:2006pb}, 
generalized to $p=0$ and arbitrary $D=2n+1$ 
for a black hole on the orbifold 
${\mathbb C}^n/{\mathbb Z}_n$ \cite{Nitta:2020pzo}.
On the other hand, 
the case of $p=3$ and $D=10$ reduces 
to a D3-brane on ${\mathbb C}^3/{\mathbb Z}_3$ 
in type-II string theory 
having near horizon geometry 
AdS${}_{5}\times S^{5}/{\mathbb Z}_3$ 
\cite{Gukov:1998kn,Oh:2002sv,Krishnan:2008kv},
and 
the case of $p=2$ and $D=11$ reduces 
to an M2-brane on ${\mathbb C}^4/{\mathbb Z}_4$ 
in M-theory 
having near horizon geometry 
AdS${}_{4}\times S^{7}/{\mathbb Z}_4$ 
  \cite{Singh:2008ix,Krishnan:2009nw}.

\subsection{The mass and charge of the black $p$-brane}

We discuss the mass and charge of the $p$-brane 
giving rise to a metric like (\ref{p:metric:Eq}).
The mass of the $p$-brane can be simply calculated 
using the ADM formalism \cite{Arnowitt:1960es, 
Arnowitt:1962hi}. 
This has already been constructed for $p$-branes in 
\cite{Argurio:1998cp, Lu:1993vt}. 
Here we use the ADM formula in our case, and 
derive the expression for the ADM mass
\Eq{
M=
\frac{V_p\,V_{{S}^{2n-1}/{\mathbb Z}_n}}{\kappa^2}
\frac{c_1}{c_0}(D-p-3)\left(
\frac{p}{p+1}c_0^{-\frac{1}{p+1}}
+\frac{D-p-2}{D-p-3}\,c_0^{\frac{1}{D-p-3}}
\right)\,,
}
where $V_p$ is the volume of the ``longitudinal'' 
space spanned by the world volume of the $p$-brane 
and $V_{S^{2n-1}/{\mathbb Z}_n}$ is the volume 
of $S^{2n-1}/{\mathbb Z}_n$. 

The charge density per a unit $p$-volume is defined by 
\Eq{
q=\frac{1}{2\kappa^2}\int_{{S}^{2n-1}/{\mathbb Z}_n}
\ast F_{(p+2)}\,.
}
Then, the total amount of the charge carried by 
the solution becomes 
\Eqr{
|Q|&=&V_p\,|q|=V_p\sqrt{\frac{2(D-2)(D-p-3)}{p+1}}\,
\frac{c_1\,V_{{S}^{2n-1}/{\mathbb Z}_n}}{2\kappa^2}
\nn\\
&=&c_0\,\sqrt{\frac{D-2}{2(p+1)(D-p-3)}}\left(
\frac{p}{p+1}c_0^{-\frac{1}{p+1}}
+\frac{D-p-2}{D-p-3}\,c_0^{\frac{1}{D-p-3}}\right)^{-1}
M\,.
}
Since the mass $M$ and the charge $Q$ are proportional 
to each other, this is extremal or BPS.

A black $p$-brane in flat space ${\mathbb C}^n$ 
has a ADM mass and charge density, given by
\Eq{
M_1=\frac{V_p\,V_{{S}^{2n-1}}}{\kappa^2}
\frac{c_1}{c_0}(D-p-3)\left(
\frac{p}{p+1}c_0^{-\frac{1}{p+1}}
+\frac{D-p-2}{D-p-3}\,c_0^{\frac{1}{D-p-3}}
\right)\,,
}
and
\Eqr{
|Q_1|&=&V_p\,|q_1|=V_p\sqrt{\frac{2(D-2)(D-p-3)}{p+1}}\,
\frac{c_1\,V_{{S}^{2n-1}}}{2\kappa^2}\nn\\
&=&c_0\,\sqrt{\frac{D-2}{2(p+1)(D-p-3)}}\left(
\frac{p}{p+1}c_0^{-\frac{1}{p+1}}
+\frac{D-p-2}{D-p-3}\,c_0^{\frac{1}{D-p-3}}\right)^{-1}
M_1\,,
}
respectively. Here we define the charge density per a 
unit $p$-volume:
\Eq{
q_1=\frac{1}{2\kappa^2}\int_{{S}^{2n-1}}\ast F_{(p+2)}\,.
} 
This black $p$-brane can live also on the 
orbifold as well. 
The geometry is not significantly modified   
if it is far away from the orbifold singularity 
where the geometry is asymptotically flat. 

Since the volumes are
$V_{{S}^{2n-1}/{\mathbb Z}_n}\, = (1/n) V_{{S}^{2n-1}}$,
our solution has $1/n$ mass and charge densities 
of the conventional one, 
\begin{equation}
M= \frac{1}{n} M_1\,, \quad |Q|= \frac{1}{n}  |Q_1|\,,
\end{equation} 
and thus we call it a {\it fractional} $p$-brane.
The unit $p$-brane $(M_1\,, Q_1)$, 
which is supposed to be minimally quantized by the 
Dirac quantization condition, can exist 
outside the orbifold singularity of ${\mathbb C}^n/{\mathbb Z}_n$,
while fractional one cannot exist and is stacked at the singularity 
for the consistency with the Dirac quantization condition. 
Such fractional objects stucked at orbifold singularities 
are common among 
Yang-Mills instantons \cite{Nakajima,Nakajima2}, 
vortices \cite{Kimura:2011wh}, 
and D-branes 
\cite{Douglas:1996sw,Douglas:1997de,Eto:2004vy}.

\section{Discussions}
  \label{sec:discussions}

We conclude with some comments on the properties 
of the solutions we have found, and potential applications. 
It is interesting to discuss the supersymmetry of 
the extremal $p$-brane solutions. It was pointed out in 
Refs.~\cite{Dabholkar:1989jt, Dabholkar:1990yf, 
Stelle:1996tz, Argurio:1998cp} that the extremal $p$-brane 
solutions are supersymmetric. 
While we have not considered supersymmetry transformation 
laws, it appears likely that some extremally charged 
black $p$-branes on the orbifolds are supersymmetric. 
This implies the existence of new varieties of solutions 
for black holes on orbifolds. 

We have discussed extended black hole solutions in 
$D$ dimensions, since this is the interest for 
supergravity. One can clearly modify the derivation 
in section \ref{sec:p} to obtain charged black $p$-branes 
with a scalar field in any dimension. It is an interesting 
question whether or not there exist black $p$-branes 
in $D$ dimensions if we add non-trivial scalar field. 
Although there exist spacetimes with this causal 
structure (taking the product of $(p+2)$-dimensional spacetime 
and $(D-p-2)$-dimensional space in the $D$-dimensional 
solution), the associated energy momentum tensor is usually 
unphysical due to giving rise to naked singularity in the 
background \cite{Argurio:1998cp}. 

There are many open questions regarding the dynamics of 
black $p$-branes on the orbifolds. Even more interesting, 
studying a way in which the classical solutions interact, 
for example by constructing solutions with intersecting 
$p$-branes \cite{Argurio:1997gt}, 
is relevant for checking the consistency with 
the interactions of these objects. Another important issue 
involving intersections of $p$-branes on the orbifolds is 
looking for supersymmetric extremal black holes with a 
non-vanishing horizon area. 
When one can identify these $p$-brane configurations 
in supergravity, mainly systems involving M-brane and 
D-branes \cite{Polchinski:1996na}, one can obtain a 
microscopic counting of states definitely giving the 
semiclassical black hole entropy \cite{Strominger:1996sh, 
Callan:1996dv}.

In Sec.~\ref{sec:p}, we have constructed 
general solutions whose transverse directions 
are general Ricci-flat manifolds. 
Thus, replacing the orbifold geometries by 
more general Ricci-flat manifolds can give 
further black holes and black $p$-brane solutions.
Recalling a fact that the resolved orbifolds
${\mathbb C}^n/{\mathbb Z}_n$ are 
complex line bundles over ${\mathbb C}{\rm P}^{n-1}$, 
one can consider Ricci-flat metrics on complex 
line bundles over other homogeneous K\"{a}hler manifolds $G/H$ 
\cite{Higashijima:2001vk,Higashijima:2001fp,Higashijima:2002px}. 
In fact, D3-branes on six-dimensional 
resolved and deformed conifolds were constructed in 
Refs.~\cite{PandoZayas:2000ctr,PandoZayas:2001iw}. 
Black $p$-branes on higher dimensional deformed conifold
\cite{Higashijima:2001yn} should be possible. 
Also, by replacing ${\mathbb C}{\rm P}^{n-1}$ in this paper by  
the quadric surface 
$G/H=$SO($n$)/[SO($n-2)\times$ U(1)] 
\cite{Higashijima:2001de},  
one could construct black $p$-branes on 
conifolds for which a horizon would be 
[SO($n$)/SO($n-2$)]/${\mathbb Z}_{n-2}$.
For more general homogeneous K\"{a}hler manifolds $G/H$, 
one can expect more general geometries with exotic horizons.

\acknowledgments

The work of M.N. is supported in part by Grant-in-Aid for Scientific
Research, JSPS KAKENHI Grant Number
JP18H01217. 
The work of K. U. is supported by Grants-in-Aid from the Scientific 
Research Fund of the Japan Society for the Promotion of 
Science, under Contract No. 16K05364. 
This work was supported by the Ministry of Education, Culture,
Sports, Science (MEXT-)Supported Program for the
Strategic Research Foundation at Private Universities
``Topological Science'' (Grant No. S1511006).


\begin{thebibliography}{99}

\bibitem{Gibbons:1987ps}
G.~W.~Gibbons and K.~i.~Maeda,
``Black Holes and Membranes in Higher Dimensional 
Theories with Dilaton Fields'', 
Nucl. Phys. B \textbf{298} (1988), 741-775.

\bibitem{Dabholkar:1989jt}
  A.~Dabholkar and J.~A.~Harvey,
  ``Nonrenormalization of the Superstring Tension'', 
  Phys.\ Rev.\ Lett.\  {\bf 63} (1989) 478.

\bibitem{Dabholkar:1990yf}
  A.~Dabholkar, G.~W.~Gibbons, J.~A.~Harvey and F.~Ruiz Ruiz,
  ``Superstrings and Solitons'', 
  Nucl.\ Phys.\ B {\bf 340} (1990) 33.

\bibitem{Callan:1991ky}
  C.~G.~Callan, Jr., J.~A.~Harvey and A.~Strominger,
  ``Worldbrane actions for string solitons'', 
  Nucl.\ Phys.\ B {\bf 367} (1991) 60.

\bibitem{Horowitz:1991cd}
G.~T.~Horowitz and A.~Strominger,
``Black strings and $p$-branes'', 
Nucl. Phys. B \textbf{360} (1991), 197-209.

\bibitem{Stelle:1996tz}
K.~S.~Stelle,
``Lectures on supergravity $p$-branes'', 
[arXiv:hep-th/9701088 [hep-th]].

\bibitem{Duff:1996hp}
M.~J.~Duff, H.~Lu and C.~N.~Pope,
``The Black branes of M theory'', 
Phys. Lett. B \textbf{382} (1996), 73-80
[arXiv:hep-th/9604052 [hep-th]].

\bibitem{Argurio:1998cp}
  R.~Argurio,
  ``Brane physics in M theory'', 
  hep-th/9807171. 

\bibitem{Bogomolny:1975de}
  E.~B.~Bogomolny,
  ``Stability of Classical Solutions'', 
  Sov.\ J.\ Nucl.\ Phys.\  {\bf 24} (1976) 449
   [Yad.\ Fiz.\  {\bf 24} (1976) 861].

\bibitem{Prasad:1975kr}
  M.~K.~Prasad and C.~M.~Sommerfield,
  ``An Exact Classical Solution for the 't Hooft Monopole and the Julia-Zee Dyon'', 
  Phys.\ Rev.\ Lett.\  {\bf 35} (1975) 760.

\bibitem{Coleman:1976uk}
  S.~R.~Coleman, S.~J.~Parke, A.~Neveu and C.~M.~Sommerfield,
  ``Can One Dent a Dyon?'', 
  Phys.\ Rev.\ D {\bf 15} (1977) 544.

\bibitem{Douglas:1996sw}
M.~R.~Douglas and G.~W.~Moore,
``D-branes, quivers, and ALE instantons'', 
[arXiv:hep-th/9603167 [hep-th]].

\bibitem{Douglas:1997de}
M.~R.~Douglas, B.~R.~Greene and D.~R.~Morrison,
``Orbifold resolution by D-branes'', 
Nucl. Phys. B \textbf{506}, 84-106 (1997)
[arXiv:hep-th/9704151 [hep-th]].

\bibitem{Gukov:1998kn}
S.~Gukov, M.~Rangamani and E.~Witten,
``Dibaryons, strings and branes in AdS orbifold models'', 
JHEP \textbf{12}, 025 (1998)
[arXiv:hep-th/9811048 [hep-th]].
  
\bibitem{Oh:2002sv}
K.~Oh and R.~Tatar,
``Orbifolds, Penrose limits and supersymmetry enhancement'', 
Phys. Rev. D \textbf{67}, 026001 (2003)
[arXiv:hep-th/0205067 [hep-th]].

\bibitem{Krishnan:2008kv}
C.~Krishnan and S.~Kuperstein,
``Gauge Theory RG Flows from a Warped Resolved Orbifold'', 
JHEP \textbf{04}, 009 (2008)
[arXiv:0801.1053 [hep-th]].

\bibitem{Singh:2008ix}
H.~Singh,
``M2-branes on a resolved $C_{4} / Z_{4}$'', 
JHEP \textbf{09}, 071 (2008)
[arXiv:0807.5016 [hep-th]].

\bibitem{Krishnan:2009nw}
C.~Krishnan, C.~Maccaferri and H.~Singh,
``M2-brane Flows and the Chern-Simons Level'', 
JHEP \textbf{05}, 114 (2009)
[arXiv:0902.0290 [hep-th]].

\bibitem{Nakajima}
H.~Nakajima, 
``Moduli spaces of anti-self-dual connections on 
ALE gravitational instantons'', 
Invent.\ Math.\ {\bf 102} (1990) 267

\bibitem{Nakajima2}
P.~B.~Kronheimer and H.~Nakajima, 
``Yang-Mills instantons on ALE gravitational instantons'', 
Math.\ Ann.\ {\bf 288} (1990) 263.

\bibitem{Kimura:2011wh}
T.~Kimura and M.~Nitta,
``Vortices on Orbifolds'', 
JHEP \textbf{09}, 118 (2011)
[arXiv:1108.3563 [hep-th]].

\bibitem{Eto:2004vy}
M.~Eto, Y.~Isozumi, M.~Nitta, K.~Ohashi, K.~Ohta and N.~Sakai,
``D-brane construction for non-Abelian walls'', 
Phys. Rev. D \textbf{71}, 125006 (2005)
[arXiv:hep-th/0412024 [hep-th]].

\bibitem{Randall:1999ee}
L.~Randall and R.~Sundrum,
``A Large mass hierarchy from a small extra dimension'', 
Phys. Rev. Lett. \textbf{83} (1999), 3370-3373
[arXiv:hep-ph/9905221 [hep-ph]].

\bibitem{Randall:1999vf}
L.~Randall and R.~Sundrum,
``An Alternative to compactification'', 
Phys. Rev. Lett. \textbf{83} (1999), 4690-4693
[arXiv:hep-th/9906064 [hep-th]].

\bibitem{Lukas:1998yy}
A.~Lukas, B.~A.~Ovrut, K.~S.~Stelle and D.~Waldram,
``The Universe as a domain wall'', 
Phys. Rev. D \textbf{59} (1999), 086001
[arXiv:hep-th/9803235 [hep-th]].

\bibitem{Lukas:1998tt}
A.~Lukas, B.~A.~Ovrut, K.~S.~Stelle and D.~Waldram,
``Heterotic M theory in five-dimensions'', 
Nucl. Phys. B \textbf{552} (1999), 246-290
[arXiv:hep-th/9806051 [hep-th]].

\bibitem{Duff:2000az}
M.~J.~Duff, J.~T.~Liu and K.~S.~Stelle,
``A Supersymmetric type IIB Randall-Sundrum realization'', 
J. Math. Phys. \textbf{42} (2001), 3027-3047
[arXiv:hep-th/0007120 [hep-th]].

\bibitem{Cvetic:2000gj}
M.~Cvetic, H.~Lu and C.~N.~Pope,
``Brane world Kaluza-Klein reductions and branes on the brane'', 
J. Math. Phys. \textbf{42} (2001), 3048-3070
[arXiv:hep-th/0009183 [hep-th]].

\bibitem{Stelle:1998xg}
  K.~S.~Stelle,
  ``BPS branes in supergravity'', 
{\it in ICTP Summer School in High-energy Physics and
Cosmology, 3, 1998}
  [hep-th/9803116].

\bibitem{Lu:1993vt}
  J.~X.~Lu,
  ``ADM masses for black strings and $p$-branes'', 
  Phys.\ Lett.\ B {\bf 313} (1993) 29
  [hep-th/9304159].

\bibitem{Arnowitt:1960es}
  R.~L.~Arnowitt, S.~Deser and C.~W.~Misner,
  ``Canonical variables for general relativity'', 
  Phys.\ Rev.\  {\bf 117} (1960) 1595.

\bibitem{Arnowitt:1962hi}
  R.~L.~Arnowitt, S.~Deser and C.~W.~Misner,
  ``The Dynamics of general relativity'', 
  Gen.\ Rel.\ Grav.\  {\bf 40} (2008) 1997
  [gr-qc/0405109].

\bibitem{Nitta:2020pzo}
  M.~Nitta and K.~Uzawa,
  ``Orbifold black holes'', 
  arXiv:2011.13316 [hep-th].

\bibitem{Ishihara:2006pb}
  H.~Ishihara, M.~Kimura, K.~Matsuno and S.~Tomizawa,
  ``Black Holes on Eguchi-Hanson Space in 
Five-Dimensional Einstein-Maxwell Theory'', 
  Phys.\ Rev.\ D {\bf 74} (2006) 047501
  [hep-th/0607035].

\bibitem{Ishihara:2006iv}
  H.~Ishihara, M.~Kimura, K.~Matsuno and S.~Tomizawa,
  ``Kaluza-Klein Multi-Black Holes in Five-Dimensional 
  Einstein-Maxwell Theory'', 
  Class.\ Quant.\ Grav.\  {\bf 23} (2006) 6919
  [hep-th/0605030].

\bibitem{Tatsuoka:2011tx}
T.~Tatsuoka, H.~Ishihara, M.~Kimura and K.~Matsuno,
``Extremal Charged Black Holes with a Twisted Extra Dimension'', 
Phys. Rev. D \textbf{85} (2012), 044006
[arXiv:1110.6731 [hep-th]].

\bibitem{Dehghani:2005zm}
  M.~H.~Dehghani and R.~B.~Mann,
  ``NUT-charged black holes in Gauss-Bonnet gravity'', 
  Phys.\ Rev.\ D {\bf 72} (2005) 124006
  [hep-th/0510083].

\bibitem{Dehghani:2006aa}
  M.~H.~Dehghani and S.~H.~Hendi,
  ``Taub-NUT/bolt black holes in Gauss-Bonnet-Maxwell gravity'', 
  Phys.\ Rev.\ D {\bf 73} (2006) 084021
  [hep-th/0602069].

\bibitem{Higashijima:2001vk}
  K.~Higashijima, T.~Kimura and M.~Nitta,
  ``Ricci flat Kahler manifolds from supersymmetric gauge theories'', 
  Nucl.\ Phys.\ B {\bf 623} (2002) 133
  [hep-th/0108084].

\bibitem{Higashijima:2001fp}
K.~Higashijima, T.~Kimura and M.~Nitta,
``Gauge theoretical construction of noncompact Calabi-Yau manifolds'', 
Annals Phys. \textbf{296}, 347-370 (2002)
[arXiv:hep-th/0110216 [hep-th]].

\bibitem{Higashijima:2002px}
  K.~Higashijima, T.~Kimura and M.~Nitta,
  ``Calabi-Yau manifolds of cohomogeneity one as complex line bundles'', 
  Nucl.\ Phys.\ B {\bf 645} (2002) 438
  [hep-th/0202064].

\bibitem{Duff:1994an}
  M.~J.~Duff, R.~R.~Khuri and J.~X.~Lu,
  ``String solitons'', 
  Phys.\ Rept.\  {\bf 259} (1995) 213
  [hep-th/9412184].

\bibitem{Tseytlin:1996hi}
  A.~A.~Tseytlin,
  ``'No force' condition and BPS combinations of $p$-branes 
in eleven-dimensions and ten-dimensions'', 
  Nucl.\ Phys.\ B {\bf 487} (1997) 141
  [hep-th/9609212].

\bibitem{Gibbons:1994vm}
G.~W.~Gibbons, G.~T.~Horowitz and P.~K.~Townsend,
``Higher dimensional resolution of dilatonic black hole 
singularities'', 
Class. Quant. Grav. \textbf{12} (1995), 297-318
[arXiv:hep-th/9410073 [hep-th]].

\bibitem{Freund:1980xh}
  P.~G.~O.~Freund and M.~A.~Rubin,
  ``Dynamics of Dimensional Reduction'', 
  Phys.\ Lett.\  {\bf 97B} (1980) 233.

\bibitem{Argurio:1997gt}
  R.~Argurio, F.~Englert and L.~Houart,
  ``Intersection rules for $p$-branes'', 
  Phys.\ Lett.\ B {\bf 398} (1997) 61
  [hep-th/9701042].

\bibitem{Polchinski:1996na}
J.~Polchinski,
``Tasi lectures on D-branes'', 
[arXiv:hep-th/9611050 [hep-th]].

\bibitem{Strominger:1996sh}
A.~Strominger and C.~Vafa,
``Microscopic origin of the Bekenstein-Hawking entropy'', 
Phys. Lett. B \textbf{379} (1996), 99-104
[arXiv:hep-th/9601029 [hep-th]].

\bibitem{Callan:1996dv}
C.~G.~Callan and J.~M.~Maldacena,
``D-brane approach to black hole quantum mechanics'', 
Nucl. Phys. B \textbf{472} (1996), 591-610
[arXiv:hep-th/9602043 [hep-th]].

\bibitem{PandoZayas:2000ctr}
L.~A.~Pando Zayas and A.~A.~Tseytlin,
``3-branes on resolved conifold'', 
JHEP \textbf{11}, 028 (2000)
[arXiv:hep-th/0010088 [hep-th]].

\bibitem{PandoZayas:2001iw}
L.~A.~Pando Zayas and A.~A.~Tseytlin,
``3-branes on spaces with $R\times S^2\times S^3$ topology'', 
Phys. Rev. D \textbf{63}, 086006 (2001)
[arXiv:hep-th/0101043 [hep-th]].


\bibitem{Higashijima:2001yn}
K.~Higashijima, T.~Kimura and M.~Nitta,
``Supersymmetric nonlinear sigma models on Ricci flat Kahler manifolds with O(N) symmetry'', 
Phys. Lett. B \textbf{515}, 421-425 (2001)
[arXiv:hep-th/0104184 [hep-th]].

\bibitem{Higashijima:2001de}
  K.~Higashijima, T.~Kimura and M.~Nitta,
  ``A Note on conifolds'', 
  Phys.\ Lett.\ B {\bf 518} (2001) 301
  [hep-th/0107100].

\end{thebibliography}
\end{document}